\documentclass[aps,preprint,showpacs,superscriptaddress,nofootinbib,amsmath,amssymb,floats,floatfix,showkeys,longbibliography]{revtex4-1}
\usepackage{comment}
\usepackage{xcolor}
\usepackage[english]{babel} 
\usepackage{graphicx}
\usepackage{subfigure}
\usepackage{palatino}
\usepackage[commandnameprefix=always]{changes}
\usepackage{hyperref}
\hypersetup{colorlinks=true,linkcolor=blue,urlcolor=blue,citecolor=blue}
\usepackage[toc,page]{appendix}
\usepackage[normalem]{ulem}
\usepackage{orcidlink}
\usepackage{lipsum}
\usepackage{graphicx}
\usepackage{subfigure}
\usepackage{palatino}
\usepackage{sans}
\usepackage{adjustbox}
\usepackage{latexsym}
\usepackage{amsmath}
\usepackage{amssymb}
\usepackage{amsfonts}
\usepackage{dcolumn}
\usepackage{bm}
\usepackage{tikz}
\usepackage{bigints}
\usepackage{array,tabularx,multirow,booktabs}
\usepackage[tracking=true]{microtype}
\SetTracking{}{500}
\SetTracking{encoding={*}, shape=sc}{40}
\UseRawInputEncoding 
\allowdisplaybreaks
\usepackage{adjustbox}
\usepackage{latexsym}
\usepackage{amsmath}
\usepackage{amssymb}
\usepackage{amsfonts}
\usepackage{dcolumn}
\usepackage{bm}
\usepackage{tikz}
\usepackage{bigints}
\usepackage{array,tabularx,multirow,booktabs}
\usepackage[tracking=true]{microtype}
\usepackage{color}
\UseRawInputEncoding 
\allowdisplaybreaks

\begin{document} 

\title{Gravitational Collapse and Formation of Regular Black Holes: Dymnikova, Hayward, and Beyond}

\author{Vitalii Vertogradov}
\email{vdvertogradov@gmail.com}
\affiliation{Physics department, Herzen state Pedagogical University of Russia,
48 Moika Emb., Saint Petersburg 191186, Russia} 
\affiliation{Center for Theoretical Physics, Khazar University, 41 Mehseti Street, Baku, AZ-1096, Azerbaijan.}
\affiliation{SPB branch of SAO RAS, 65 Pulkovskoe Rd, Saint Petersburg
196140, Russia}
\begin{abstract}
The gravitational collapse of a star can lead to the formation of a regular black hole. However, a key factor in this process is the transition of ordinary baryonic matter into a substance that forms the de Sitter core.  However, the formation of de Sitter core during gravitational collapse remains an open question, particularly since ordinary baryonic matter does not naturally transition into the exotic matter required to form a de Sitter core. In this paper, we investigate the gravitational collapse of baryonic matter and its potential to form well-known regular black hole solutions, such as those proposed by Dymnikova and Hayward. We model the collapse process as a transition of baryonic matter into a new type of matter, accompanied by the release of energy in the form of electromagnetic radiation. Using a generalized dynamical framework, we derive the energy density of the emitted radiation as a function of both the properties of the initial baryonic matter and the resulting exotic matter. Our findings demonstrate that the gravitational collapse can lead to the formation of various types of regular black holes, providing insights into the physical mechanisms underlying their creation. The detectable radiation signature offers a potential observational test for distinguishing between different black hole models.
\end{abstract}

\date{\today}



\maketitle

\section{Introduction}
In 2019, the Event Horizon Telescope collaboration published the first images of black hole shadows at the center of the M87 galaxy~\cite{bib:ehtm1}. This breakthrough made the study of black hole shadow properties a key tool for determining which black hole models are valid and which should be discarded. Additionally, investigating shadow properties can help identify viable modifications to general relativity and rule out those inconsistent with experimental data~\cite{bib:vanozi}.

However, when discussing black holes, it is essential to consider that they possess not only an event horizon but also a central singularity. While the singularity at the event horizon is purely coordinate-based and can be eliminated by an appropriate choice of coordinates, the central singularity is intrinsic and cannot be removed by any coordinate transformation. The presence of a singularity indicates that the underlying theory is no longer applicable to describe such objects. Consequently, significant attention has recently been given to black holes without a central singularity—so-called regular black holes. One of the earliest ideas, proposed by Gliner~\cite{bib:gliner} and Sakharov~\cite{bib:sakharov}, suggested that matter at high densities transitions into a vacuum-like state described by the de Sitter metric. This concept was later developed in the works of Dymnikova~\cite{bib:dym}. Bardeen was the first to construct a model of a regular black hole, choosing a mass function $M(r)$ to ensure the finiteness of curvature invariants~\cite{bib:bardeen}—such as the Kretschmann scalar, the square of the Ricci tensor, and the Ricci scalar—throughout spacetime. However, a significant drawback of such phenomenological approaches is the appearance of an energy-momentum tensor in the right-hand side of Einstein's equations, whose physical nature often remains unclear. Subsequently, it was realized that nonlinear electrodynamics could serve as the source for Bardeen's black hole~\cite{bib:garcia}, and Bronnikov~\cite{bib:bronnikov} proved that if nonlinear electrodynamics generates a regular black hole, it must describe a magnetic monopole with no electric field.

Further developments in the theory led to numerous models of regular black holes~\cite{bib:rincon1, bib:rincon2, bib:ingeniring, bib:hay, bib:ali2025cqg, bib:ali2025brazil, bib:daniil2025barionic, bib:khlopov, bib:khlopov2, bib:khlopov3, bib:khlopov4, bib:khlopov5, bib:baptista2024}, many of which are purely mathematical solutions with little connection to reality(See comprehensive review~\cite{bib:19, bib:jap}). Moreover, a regular core implies that the matter forming the de Sitter core must violate the strong energy conditions, which state that gravity is attractive. Furthermore, most solutions are limited to describing de Sitter cores and cannot serve as models for testing experimental manifestations of black holes through shadow observations. This raises the question: what lies beneath the event horizon? Is there a singularity or a de Sitter core? An even more critical question regarding regular black holes is that ordinary massive stars lack exotic matter, which prevents the formation of a regular core—a de Sitter core. Therefore, it is necessary to study gravitational collapse processes in massive stars and examine the behavior of matter at critical densities.

Thus, the simplest way to determine whether a regular black hole forms is to observe the gravitational collapse of a massive star. In \cite{bib:vertogradov2025podu}, it was demonstrated that the gravitational collapse of dust transitioning into radiation can lead to the formation of a regular black hole. A recent study \cite{bib:ali2025new} showed that a regular black hole may form during the gravitational collapse of baryonic matter, which transitions into quark-gluon plasma at critical densities.

In this article, we investigate the gravitational collapse of baryonic matter and explore whether such a process can lead to known solutions of Einstein's equations describing regular black holes, such as the Hayward, Dymnikova, and other metrics.

We analyze the gravitational collapse of baryonic matter described by the equation of state $P = \alpha \rho$. This model is represented by the Husain black hole~\cite{bib:husain1996exact}. At critical densities, baryonic matter transitions into a new type of matter with some efficiency $\beta$, which generally depends on the radial coordinate and enhances the effect near the center. During this transition, energy is released in the form of electromagnetic radiation. Both the conversion efficiency and the radiation energy density directly depend on the properties of the baryonic matter and the newly formed substance, allowing us to infer the nature of the matter formed during gravitational collapse and whether the final outcome will be a regular black hole after horizons form.

We chose the Husain solution as the simplest dynamic black hole solution of Einstein's equations. However, the equation of state $P = \alpha \rho$ does not generally describe baryonic matter. Nevertheless, generalization to baryonic matter is straightforward by redefining the equation-of-state parameter $\alpha$. The most physically realistic equation of state for baryonic matter is:
\begin{eqnarray}
\bar{P} = \frac{1}{3}(P_r + 2P_t), \nonumber \\
\bar{P} = \omega \rho.
\end{eqnarray}
The generalization to this case is direct. To achieve this, we set:
\begin{equation}
\alpha = \frac{1}{2}(3\omega + 1).
\end{equation}
Instead of the Husain solution, we obtain a dynamic generalization of the Kiselev solution~\cite{bib:kiselev, bib:tur1, bib:tur2, bib:tur3}. However, for computational simplicity, it is convenient to use the parameter $\alpha$, as adopted in this model. It is important to note that the results remain valid and can be generalized directly to the parameter $\omega$. Moreover, in the case of $\alpha =1$ ($\omega=\frac{1}{3}$) the properties of black hole shadow formation in dynamical case has been investigated~\cite{bib:yaghoub2024epjc}.

The article is organized as follows. In Sec. 2, we discuss the general characteristics of regular black holes. In Sec. 3, we describe a model of gravitational collapse in which baryonic matter transitions into radiation and forms a new type of matter. In Sec. 4, the developed method is applied to describe processes leading to the formation of Hayward and Dymnikova regular black holes. In Sec. 5, we generalize the method to arbitrary types of matter and arbitrary black holes, deriving the radiation energy density as a function of the parameters of baryonic and new matter. Sec. 6 provides the conclusion.

This article uses the geometrized unit system where $c = 8\pi G = 1$ and the signature $-+++$ is adopted. Primes and dots denote differentiation with respect to the radial $r$ and temporal $v$ coordinates, respectively.

\section{Regular Black Hole Description}

In this section, we briefly describe regular black holes. 
Consider a spherically symmetric dynamic metric describing a black hole in the form:
\begin{equation} \label{eq:metric}
ds^2 = -f(v,r)dv^2 + 2dvdr + r^2d\Omega^2,
\end{equation}
where
\begin{equation}
f(v,r) = 1 - \frac{2M(v,r)}{r},
\end{equation}
and $M(v,r)$ is the mass function depending on time $v$ and radial coordinate $r$. Here, $d\Omega^2 = d\theta^2 + \sin^2\theta d\varphi^2$ represents the metric on the unit 2-sphere.

The spacetime described by \eqref{eq:metric} corresponds to the following matter distribution:
\begin{eqnarray} \label{eq:dengen}
\sigma &=& 2\frac{\dot{M}}{r^2}, \nonumber \\
\rho &=& \frac{2M'}{r^2}, \nonumber \\
P &=& -\frac{M''}{r},
\end{eqnarray}
where $\sigma$ is the energy flux density, while $\rho$ and $P$ are the energy density and pressure of the matter, respectively.

Furthermore, the conservation law for the energy-momentum tensor $T^{ik}_{;k} = 0$ leads to an equation that we will refer to as the continuity equation:
\begin{equation}
\rho'r + 2\rho + 2P = 0.
\end{equation}
Note that this equation can also be derived by taking the derivative of the energy density $\rho$ from \eqref{eq:dengen}.

To ensure that the black hole is regular throughout spacetime, we calculate the curvature invariants: the Ricci scalar $R$, the square of the Ricci tensor $S = R_{ik}R^{ik}$, and the Kretschmann scalar $K = R_{iklm}R^{iklm}$. In terms of the mass function $M(v,r)$, these invariants take the form:
\begin{eqnarray} \label{eq:curvature}
R &=& \frac{4M' + 2rM''}{r^2}, \nonumber \\
S &=& \frac{8M'^2 + 2r^2M''^2}{r^4}, \nonumber \\
K &=& \frac{48M^2 - 64rMM' + 32r^2M'^2 + 16r^2MM'' - 16r^3M'M'' + 4r^4M''^2}{r^6}.
\end{eqnarray}

Alternatively, these curvature invariants can be expressed in terms of the mass function $M(v,r)$, energy density $\rho$, and pressure $P$ from \eqref{eq:dengen} as follows:
\begin{eqnarray} \label{eq:curvature2}
R &=& 2\rho - 2P, \nonumber \\
S &=& 2\rho^2 + 2P^2, \nonumber \\
K &=& \frac{48M^2}{r^6} - \frac{16M}{r^3}(2\rho - P) + 8\rho^2 - 8\rho P + 4P^2.
\end{eqnarray}

Thus, the statement that a black hole contains no singularities can be characterized by three conditions:
\begin{enumerate}
\item The energy density $\rho$ must remain finite throughout spacetime. Specifically, the following condition must hold:
\begin{equation}
\lim\limits_{r \to 0} \rho = \rho_0.
\end{equation}
\item The matter pressure must be finite throughout spacetime, satisfying:
\begin{equation}
\lim\limits_{r \to 0} P = P_0.
\end{equation}
\item The mass function must vanish at the center, i.e.,
\begin{equation}
\lim\limits_{r \to 0} M(v,r) = 0.
\end{equation}
\end{enumerate}

A few remarks are in order regarding these three conditions. First, we say that a regular black hole possesses a de Sitter core if the following condition is satisfied:
\begin{equation}
\lim\limits_{r \to 0} P = -\lim\limits_{r \to 0} \rho.
\end{equation}
Second, there is no need to impose additional constraints on the first and second derivatives of the mass function, as they are already accounted for by the finiteness of the energy density and pressure. Thus, it suffices to mention that the mass function vanishes at the center.

Dymnikova proposed constructing solutions for regular black holes based on an energy density of the form~\cite{bib:dym}:
\begin{equation}
\rho = \varepsilon e^{-\frac{r^3}{2Mr_0^2}},
\end{equation}
where $\varepsilon$ and $r_0$ are related by the de Sitter relation:
\begin{equation}
\varepsilon_0 = \frac{3}{r_0^2}.
\end{equation}
In this case, solving Einstein's equations leads to the spacetime metric known as the Dymnikova black hole:
\begin{equation}
ds^2 = -fdv^2 + 2dvdr + r^2d\Omega^2,
\end{equation}
where
\begin{equation}
f(r) = 1 - \frac{2M}{r}\left(1 - e^{-\frac{r^3}{2Mr_0^2}}\right).
\end{equation}

Another well-known solution was obtained by Hayward~\cite{bib:hay} in a study of the formation and evaporation of regular black holes. This metric is not constructed from a specific matter distribution but is instead postulated as a minimal model:
\begin{equation}
f(v,r) = 1 - \frac{2M(v)r^2}{r^3 + 2M(v)L^2}.
\end{equation}
From this, the energy density is derived as:
\begin{equation}
\rho = \frac{12M(v)^2L^2}{(r^3 + 2M(v)L^2)^2},
\end{equation}
and the pressure is given by:
\begin{equation}
P = 24M^2(v)L^2\frac{r^3 - M(v)L^2}{(r^3 + 2M(v)L^2)^3}.
\end{equation}
Initially, Hayward introduced the parameter $L$, of the order of the Planck length, as a regularization parameter. However, it was later shown that $L$ could represent a magnetic monopole charge arising from nonlinear electrodynamics, which supports the Hayward solution~\cite{bib:bronnikov}.

In this article, we do not attempt to determine the nature of the matter supporting the Dymnikova, Hayward, or other regular black hole solutions. Instead, we aim to explain how such matter could form during the collapse of baryonic matter. Furthermore, we argue that this process releases energy in the form of electromagnetic radiation, which may be detectable. Additionally, we demonstrate that for each type of matter, the energy flux density depends on both the parameters of the new matter and those of the baryonic matter, making it possible to distinguish between different black hole solutions when studying the collapse of massive stars.

\section{General Considerations}
We consider a model in which baryonic matter transitions into radiation. This implies that the total energy-momentum tensor is conserved, while its individual components are not. Specifically, we examine a model where baryonic matter is described by the equation of state
\begin{equation}
P = \alpha \rho, \quad 0 \leq \alpha \leq 1, \quad \alpha \neq \frac{1}{2},
\end{equation}
and transitions into radiation, which is governed by the equation
\begin{equation}
P = \frac{1}{3} \rho.
\end{equation}
In this scenario, the continuity equation takes the form
\begin{eqnarray} \label{eq:system}
\rho_b' r + 2P_b + 2\rho_b &=& -\beta(v,r)\rho_r, \nonumber \\
\rho_r' r + 2P_r + 2\rho_r &=& \beta(v,r)\rho_r.
\end{eqnarray}
Here, $\beta$ represents the dimensionless transition rate of baryonic matter into radiation. We impose the condition that the denser the object becomes, the faster the transition process occurs, which requires $\beta' < 0$.

Solving the second equation in the system, we obtain the radiation density as
\begin{equation} \label{eq:den_rad}
\rho_r = \rho_{0r}(v) e^{\int \frac{\beta - \frac{8}{3}}{r} dr},
\end{equation}
where $\rho_{0r}(v)$ is a positive integration function. Substituting the radiation density \eqref{eq:den_rad} into the first equation of the system \eqref{eq:system}, we find the baryonic matter density:
\begin{equation} \label{eq:den_bar}
\rho_b = r^{-(2+2\alpha)} \left[ C(v) - \rho_{0r} \int r^{1+2\alpha} \beta e^{\int \frac{\beta - \frac{8}{3}}{r} dr} dr \right].
\end{equation}

It has been demonstrated in previous works~\cite{bib:vertogradov2025podu, bib:ali2025new} that a regular black hole can be obtained with an appropriate choice of the function $\beta$. We now show that specific choices of $\beta$ lead to the solutions of Dymnikova, Hayward, and Bardeen solutions.

Before proceeding, we note that $C(v)$ in \eqref{eq:den_bar} is an integration function associated with the baryonic matter density in the absence of interaction. Since we assume interaction between matter and radiation, we set $C(v) \equiv 0$. Consequently, the total energy density of the system is given by
\begin{equation} \label{eq:den_tot}
\rho = \rho_r + \rho_b = \rho_{0r}(v)e^{\int \frac{\beta - \frac{8}{3}}{r} dr} - r^{-2\alpha-2} \rho_{0r} \int r^{1+2\alpha} \beta e^{\int \frac{\beta - \frac{8}{3}}{r} dr} dr.
\end{equation}

\subsection{Dymnikova Solution}
Based on these assumptions, we demonstrate that this framework can lead to Dymnikova's black hole solution. To achieve this, consider the function $\beta$ in the form
\begin{equation} \label{eq:betadym}
\beta(v,r) = \frac{8}{3} - \frac{3r^3}{2M(v)r_0(v)^2} \left( 1 + \frac{3}{2+2\alpha - \frac{3r^3}{2M(v)r_0^2(v)}} \right).
\end{equation}
This yields
\begin{equation}
\int \frac{\beta - \frac{8}{3}}{r} dr = \ln |2\alpha + 2 - \frac{3r^3}{2Mr_0^2}| - \frac{r^3}{2Mr_0^2}.
\end{equation}
The radiation energy density is then expressed as
\begin{equation} \label{eq:den_dym_rad}
\rho_r = \rho_{0r}(v) \left( 2+2\alpha - \frac{3r^3}{2Mr_0^2} \right) e^{-\frac{r^3}{2Mr_0^2}}.
\end{equation}

To determine the baryonic matter density, we consider a function of the form
\begin{equation}
f(r) = \left( ar^{2\alpha+2} + br^{2\alpha+5} \right) e^{-\frac{r^3}{2Mr_0^2}},
\end{equation}
and compute its derivative with respect to $r$:
\begin{equation} \label{eq:promdym}
f'(r) = \left[ a(2\alpha+2)r^{2\alpha+1} + \left( (2\alpha+5)b - \frac{3a}{2Mr_0^2} \right) r^{2\alpha+4} - \frac{3b}{2Mr_0^2} r^{2\alpha+5} \right] e^{-\frac{r^3}{2Mr_0^2}}.
\end{equation}

Next, consider the integral in the expression for the baryonic energy density \eqref{eq:den_bar}:
\begin{eqnarray} \label{eq:pd}
&& \int r^{1+2\alpha} \beta e^{-\frac{r^3}{2Mr_0^2}} dr = \int \left[ \frac{8}{3}(2\alpha+2)r^{1+2\alpha} \right. \nonumber \\
&& \left. + \left( -\frac{8}{2Mr_0^2} - (2\alpha+5)\frac{3}{2Mr_0^2} \right) r^{2\alpha+4} + \frac{9}{4M^2r_0^4} r^{2\alpha+5} \right] e^{-\frac{r^3}{2Mr_0^2}} dr.
\end{eqnarray}
By comparing coefficients of corresponding powers of $r$ in \eqref{eq:pd} and \eqref{eq:promdym}, we find
\begin{equation}
a = \frac{8}{3}, \quad b = -\frac{3}{2Mr_0^2}.
\end{equation}
Thus,
\begin{equation}
\int r^{1+2\alpha} \beta e^{-\frac{r^3}{2Mr_0^2}} dr = \left[ \frac{8}{3}r^{2\alpha+2} - \frac{3}{2Mr_0^2}r^{2\alpha+5} \right] e^{-\frac{r^3}{2Mr_0^2}}.
\end{equation}

The baryonic energy density transitioning into radiation can then be written as
\begin{equation} \label{eq:den_dym_bar}
\rho_b = \rho_{0r} \left[ \frac{3r^3}{2Mr_0^2} - \frac{8}{3} \right] e^{-\frac{r^3}{2Mr_0^2}}.
\end{equation}

The total energy density of the system, combining the radiation density \eqref{eq:den_dym_rad} and the baryonic energy density \eqref{eq:den_dym_bar}, is
\begin{equation}
\rho = \rho_r + \rho_b = \rho_{0r} \left( 2\alpha - \frac{2}{3} \right) e^{-\frac{r^3}{2Mr_0^2}}.
\end{equation}

Not all barotropic matter leads to Dymnikova's solution. Since $\rho_{0r}(v) > 0$ for all $v$, satisfying the weak energy conditions requires $\alpha > \frac{1}{3}$. From the energy density corresponding to Dymnikova's solution,
\begin{equation}
\rho_{\text{Dymnikova}} = \frac{6}{r_0^2} e^{-\frac{r^3}{2Mr_0^2}},
\end{equation}
we deduce the definition of $\rho_{0r}(v)$ as
\begin{equation}
\rho_{0r} \equiv \frac{6}{\left( 2\alpha - \frac{2}{3} \right) r_0^2}.
\end{equation}

With this definition, solving Einstein's equations yields the dynamic Dymnikova black hole metric:
\begin{equation}
ds^2 = -f(v,r) dv^2 + 2 dv dr + r^2 d\Omega^2,
\end{equation}
where
\begin{equation}
f(v,r) = 1 - \frac{2M}{r} \left( 1 - e^{-\frac{r^3}{2Mr_0^2}} \right).
\end{equation}

\subsection{Hayward Spacetime}

Let us now examine how the application of this method leads to the well-known Hayward solution. For this purpose, we choose the function $\beta$ in the form:
\begin{equation}
\beta(v,r) = \frac{\frac{2}{3}\alpha \xi(r) - \left(2\alpha + \frac{8}{3}\right)\eta(r,v) - \eta' r}{\alpha \xi - \eta},
\end{equation}
where
\begin{eqnarray} \label{eq:defxieta}
\xi &=& \frac{12M(v)^2L(v)^2}{(r^3 + 2M(v)L^2(v))^2}, \nonumber \\
\eta &=& \frac{24M(v)^2L(v)^2(r^3 - M(v)L^2(v))}{(r^3 + 2M(v)L^2(v))^3}.
\end{eqnarray}
With this definition, we obtain:
\begin{equation}
\int \frac{\beta - \frac{8}{3}}{r} dr = \ln |\alpha \xi - \eta|.
\end{equation}
Consequently, the energy density of radiation takes the form:
\begin{equation}
\rho_r = \rho_{0r}(v)(\alpha \xi - \eta).
\end{equation}

To determine the energy density of baryonic matter $\rho_b$, we introduce the function $g(r)$ as follows:
\begin{equation}
g(r) \equiv r^{2\alpha+2}\left(a \xi + b\eta\right),
\end{equation}
and consider its derivative. This yields:
\begin{equation} \label{eq:der1}
g'(r) = r^{1-2\alpha}\left[(2\alpha+2)(a \xi + b\eta) + a\xi'r + b\eta'r\right],
\end{equation}
where $a$ and $b$ are arbitrary constants.

From the definitions of $\xi$ and $\eta$, we observe the following relation:
\begin{equation}
\xi'r = -2\eta - 2\xi.
\end{equation}
Substituting this relation into \eqref{eq:der1}, we find:
\begin{equation} \label{eq:int_hay}
g'(r) = r^{1+2\alpha}\left[2\alpha a \xi + [(2\alpha+2)b - 2a]\eta + b\eta'r\right].
\end{equation}

Now, we analyze the integral in the definition of $\rho_b$. The integrand takes the form:
\begin{equation}
r^{1+2\alpha}\beta e^{\int \frac{\beta-\frac{8}{3}}{r}dr} = r^{1-2\alpha}\left[\frac{2}{3}\alpha \xi - \left(\frac{8}{3} + 2\alpha\right)\eta - \eta'r\right].
\end{equation}
By comparing the expressions in square brackets with \eqref{eq:int_hay}, we deduce that $a$ and $b$ must satisfy:
\begin{equation}
\alpha = \frac{1}{3}, \quad b = -1.
\end{equation}
Thus, we can write:
\begin{equation}
\int r^{1-2\alpha}\beta e^{\int \frac{\beta-\frac{8}{3}}{r}dr} dr = r^{2\alpha+2}\left(\frac{1}{3}\xi - \eta\right).
\end{equation}
The energy density of baryonic matter is then given by:
\begin{equation}
\rho_b = -\rho_{0r}\left(\frac{1}{3}\xi - \eta\right).
\end{equation}

The total energy density is expressed as:
\begin{equation}
\rho = \rho_r + \rho_b = \rho_{0r}\left(\alpha - \frac{1}{3}\eta\right).
\end{equation}
Using the definition of $\eta$ from \eqref{eq:defxieta} and setting $\rho_{0r} = \frac{1}{\alpha - \frac{1}{3}}$, we find:
\begin{equation}
\rho = \frac{12M(v)^2L(v)^2}{(r^3 + 2M(v)L^2(v))^2},
\end{equation}
which corresponds precisely to the energy density associated with the Hayward solution.

A few remarks are in order regarding the determination of $\rho_{0r}$. First, $\alpha = \frac{1}{3}$ is excluded because the transition from radiation to radiation is uninteresting and is described by Husain's solution for $\alpha = \frac{1}{3}$. Second, to avoid violating the weak energy conditions, we must impose $\alpha > \frac{1}{3}$. Thus, the physically motivated transition of baryonic matter into matter supporting the Hayward solution, with energy release in the form of radiation, is possible only if the equation-of-state parameter of baryonic matter satisfies $\alpha > \frac{1}{3}$.
\section{Arbitrary Black Hole Solution}

We now develop a method that not only demonstrates the possibility of explaining the formation of any black hole through a given process but also allows us to estimate the energy density of radiation emitted during this process. 

To begin, we rewrite the energy densities of radiation and baryonic matter:
\begin{eqnarray}
\rho_r &=& \rho_{0r} e^{\int \frac{\beta - \frac{8}{3}}{r} dr}, \nonumber \\
\rho_b &=& r^{-(2\alpha + 2)} \left[ C(v) - \rho_{0r} \int r^{1 + 2\alpha} \beta e^{\int \frac{\beta - \frac{8}{3}}{r} dr} dr \right].
\end{eqnarray}
Our goal is to describe the transition of baryonic matter into a new type of matter $\rho_{new}$, during which energy is released in the form of radiation with energy density $\rho_r$. Clearly, each type of matter corresponds to its own conversion efficiency $\beta$. To determine $\beta$, we solve the integral equation:
\begin{equation}
\rho_{new} = \rho_r + \rho_b.
\end{equation}
During this process, the amount of baryonic matter decreases over time, while the amount of new matter increases. Consequently, the parameter $C(v)$ decreases with time, and the complete transition of baryonic matter into the new type occurs at a time $v = v_{end}$ such that $C(v_{end}) = 0$. In our reasoning, we assume that the transition is complete, although the function $\beta$ will have the same value regardless of whether the transition is fully completed or not. The function $\beta$ simply reflects the efficiency of the process and how quickly it occurs.

The solution to the integral equation is:
\begin{equation}
\beta = \frac{\frac{2}{3} \alpha \rho_{new} - \left( 2\alpha + \frac{8}{3} \right) P_{new} - P'_{new} r}{\alpha \rho_{new} - P_{new}}.
\end{equation}
Thus,
\begin{equation}
\int \frac{\beta - \frac{8}{3}}{r} dr = \ln |\alpha \rho_{new} - P_{new}|.
\end{equation}
An important note: when evaluating the integral, we must use the continuity equation for the new type of matter, which is:
\begin{equation} \label{eq:nepr}
\rho'_{new} r + 2 \rho_{new} + 2 P_{new} = 0.
\end{equation}
Therefore, the energy density of the emitted radiation during the transition of baryonic matter into the new type depends on the parameters of the new matter and the equation-of-state parameter of baryonic matter $\alpha$.

\textbf{The key result is:}
\begin{equation}
\rho_r = \rho_{0r} \left( \alpha \rho_{new} - P_{new} \right).
\end{equation}

To determine the depletion of baryonic matter, we introduce a new function $g$:
\begin{equation}
g(r) \equiv r^{2\alpha + 2} \left( a \rho_{new} + b P_{new} \right).
\end{equation}
Taking the derivative with respect to $r$ and using the continuity equation \eqref{eq:nepr} for $\rho'_{new} r$, we find:
\begin{equation} \label{eq:gen}
g'(r) = r^{1 + 2\alpha} \left[ 2\alpha a \rho_{new} + \left( (2\alpha + 2)b - 2a \right) P_{new} + b P'_{new} r \right].
\end{equation}

Substituting $\beta$ into the integrand of $\rho_b$, we obtain:
\begin{equation}
r^{1 + 2\alpha} \beta e^{\int \frac{\beta - \frac{8}{3}}{r} dr} = r^{1 + 2\alpha} \left[ \frac{2\alpha}{3} \rho_{new} - \left( \frac{8}{3} + 2\alpha \right) P_{new} - P'_{new} r \right].
\end{equation}
Comparing coefficients with \eqref{eq:gen}, we find:
\begin{equation}
a = \frac{1}{3}, \quad b = -1.
\end{equation}

The energy density of baryonic matter is then given by:
\begin{equation}
\rho_b = -\rho_{0r} \left( \frac{1}{3} \rho_{new} - P_{new} \right).
\end{equation}

The total energy density of the new matter is:
\begin{equation}
\rho_r + \rho_b = \rho_{0r} \left( \alpha - \frac{1}{3} \right) \rho_{new}.
\end{equation}

Finally, the integration constant $\rho_{0r}$ is determined as:
\begin{equation}
\rho_{0r} = \frac{1}{\alpha - \frac{1}{3}}.
\end{equation}
Since this constant must be positive, $\alpha$ must satisfy $\alpha > \frac{1}{3}$.
\section{Conclusion}

When considering the process of gravitational stellar collapse, phase transitions of matter must be taken into account. This is because, during the later stages of collapse, the density of matter reaches critical values, leading to a transition into a new state of matter. Such processes are expected to release energy in the form of electromagnetic radiation, which can be detected by a distant observer.

In this work, we examined the collapse of baryon-like matter transitioning into a new state and estimated the energy density emitted as electromagnetic radiation. We demonstrated that such processes may lead not only to well-known solutions of Einstein's equations, such as Hayward's and Dymnikova's regular black holes but also to any other black hole whose metric is a solution of Einstein's equations.

One of the key results of this paper is the derivation of a formula linking the energy density of radiation released during the transition of baryon-like matter to a new state of matter. This formula depends on the parameters of both the new matter and the baryon-like matter. Additionally, we showed how the function $\beta$, which can be associated with the efficiency of matter-to-radiation conversion, is related to the properties of the new state of matter.

The term "baryon-like" was introduced to emphasize that the Husain solution does not correspond to conventional baryonic matter. Specifically, the equation of state $P = \alpha \rho$ differs from its counterpart in cosmology due to pressure anisotropy. However, a direct transition to classical baryonic matter can be achieved by redefining the constant $\alpha$ as follows:
\begin{equation}
    \alpha = \frac{1}{2} \left( 3\omega + 1 \right).
\end{equation}
Despite this, the model of baryon-like matter proves more convenient for calculations, and the results obtained in this paper remain valid when redefining the equation-of-state constant $\alpha$ in terms of $\omega$. The analysis of radiation energy density for various black hole models may play a pivotal role in astrophysics, providing a means to evaluate the validity of proposed black hole models.


\bibliography{ref}

\end{document}